\documentclass[10pt]{iopart}

\usepackage{bm}
\usepackage{graphicx}
\usepackage{amssymb}

\expandafter\let\csname equation*\endcsname\relax
\expandafter\let\csname endequation*\endcsname\relax
\usepackage{amsmath}

\begin{document}

\title{Tracking a hysteretic and disorder-broadened phase transition via the electromagnon response in improper ferroelectrics}

\author{C.~D.~W.~Mosley$^1$, D. Prabhakaran$^2$ and J.~Lloyd-Hughes$^1$}
\address{$^1$Department of Physics, University of Warwick, Gibbet Hill Road, Coventry CV4 7AL, UK}
\address{$^2$Department of Physics, Clarendon Laboratory, University of Oxford, Parks Road, Oxford OX1 3PU, UK}

\begin{abstract}
We demonstrate that electromagnons can be used to directly probe the nature of a phase transition between magnetically ordered phases in an improper ferroelectric. The antiferromagnetic/paraelectric to antiferromagnetic/ferroelectric phase transition in Cu$_{1-x}$Zn$_{x}$O ($x=0, 0.05$) alloys was tracked via the electromagnon response using terahertz time-domain spectroscopy, on heating and cooling through the phase transition. The transition was found to exhibit thermal hysteresis, confirming its first-order nature, and to broaden under the influence of spin-disorder upon Zn substitution. The energy of the electromagnon increases upon alloying, as a result of the non-magnetic ions modifying the magnetic interactions that give rise to the multiferroic phase and electromagnons. We describe our findings in the context of recent theoretical work that examined improper ferroelectricity and electromagnons in CuO from phenomenological and first-principles approaches.
\end{abstract}

\ioptwocol

\section*{Introduction}

Multiferroic materials have increasingly inspired interest in recent years due to their potential technological applications; specifically magnetoelectric coupling, where the magnetic order can be influenced by an applied electric field and vice versa, may have intriguing applications in spintronics \cite{Kostylev2005,Tong2016,Zanolli2016} and novel electronic components \cite{Gajek2007,Yang2010}. Of particular interest are improper ferroelectric multiferroics, in which ferroelectricity is induced by frustrated magnetic interactions, forming a magnetic state that breaks inversion symmetry, for instance an incommensurate spin-cycloid \cite{Katsura2005,Kenzelmann2005}. Strong magnetoelectric coupling can occur in such materials \cite{Kimura2003,Wang2009}, however the improper ferroelectric phase generally occurs only at low temperatures, typically below $\sim70$\,K \cite{Kimura2003}. For many of the technological applications to be realized, room-temperature magnetoelectrics are strongly desired.

The static polarization in an improper ferroelectric phase can be understood as arising from the spin current or inverse Dzyaloshinskii-Moriya (DM) interaction \cite{Katsura2005}, which induces a polarization $\mathbf{P}\propto\left(\mathbf{S}_{n}\times\mathbf{S}_{n+1}\right)$, where $\mathbf{S}_{n}$, $\mathbf{S}_{n+1}$ are adjacent spins in the cycloid. The induced static polarization in many improper ferroelectric materials is weak (typically $<1$\,$\mu$C\,cm$^{-2}$) \cite{Kimura2003}. Further, improper ferroelectrics exhibit dynamic magnetoelectric coupling, whereby an oscillating electric field couples to a spin wave, or magnon. This results in a novel quasiparticle excitation at terahertz frequencies - the electromagnon \cite{Pimenov2006,Pimenov2006a,Sushkov2007,Katsura2007,Kida2009,Jones2014}. Two distinct mechanisms that give rise to electromagnons have been discussed in the literature: Dzyaloshinskii-Moriya electromagnons, which are eigenmodes of a spin-cycloid induced by the DM interaction \cite{Katsura2007} and couple directly to oscillating electric fields \cite{DeSousa2008}; and exchange-striction electromagnons, which arise from a modulation of the isotropic Heisenberg exchange interaction proportional to a ($\mathbf{S}_{i}\cdot\mathbf{S}_{j}$) term in the Hamiltonian \cite{Sushkov2008,ValdesAguilar2009}. Electromagnons have been observed at low temperatures ($<70$\,K) in materials such as $R$MnO$_{3}$ and $R$Mn$_{2}$O$_{5}$ ($R=$ rare earths) \cite{Pimenov2006,Pimenov2006a,Sushkov2007,Katsura2007,Kida2009}, and at high temperatures in Cu$_{1-x}$Zn$_{x}$O alloys (213\,-\,230\,K in $x=0$ and 159\,-\,190\,K in $x=0.05$) \cite{Jones2014,Jones2014a}. An IR and Raman-active electromagnon has also recently been observed at up to 250K in a z-type hexaferrite \cite{Kadlec2016}.

In this paper we first review the salient features of improper ferroelectricity in pure and spin-disordered CuO. We then demonstrate that the electromagnon can be used to probe order-to-order phase transitions, such as found in CuO, where one phase is an improper ferroelectric. A detailed study of the antiferromagnetic/paraelectric to antiferromagnetic/ferroelectric phase transition using terahertz time-domain spectroscopy highlighted the hysteretic nature of the phase transition, and the influence of spin-disorder.

\section*{Improper ferroelectricity in CuO}

A promising material system in the study of improper ferroelectrics is cupric oxide (CuO), which exhibits a magnetically-induced ferroelectric phase with spin-cycloidal ordering at $\sim$230\,K \cite{Kimura2008}. CuO has a monoclinic crystal structure with space group $C2/c$, which can be visualized as zig-zagging Cu-O-Cu chains along the [101] and [10$\bar{1}$] directions. The magnetic phases of CuO have previously been characterized by neutron diffraction \cite{Forsyth1988,Ain1992} and ultrasound velocity measurements \cite{Villarreal2012}. Below 213\,K, the dominant magnetic interaction is antiferromagnetic superexchange between spins in Cu$^{2+}$ chains in the [10$\bar{1}$] direction ($J_{\mathrm{AFM}}\sim80$\,meV) \cite{Ain1989}. Weaker ferromagnetic superexchange interactions occur between spins in adjacent spin chains along the [101] ($J_{\mathrm{FM1}}\sim5$\,meV) and [010] ($J_{\mathrm{FM2}}\sim3$\,meV) directions \cite{Ain1989}. As the ratio of interchain to intrachain interactions is about 0.1, CuO can be described as a quasi-1D collinear Heisenberg antiferromagnet in the low temperature (AF1) phase \cite{Boothroyd1997,Shimizu2003}, consisting of two interpenetrating Cu$^{2+}$ sublattices with spins aligned along the $b$-axis. At 213\,K the spins on one sublattice flop into the $ac$-plane \cite{Yablonskii1990} and form an incommensurate spin-cycloid phase (AF2) with magnetic ordering vector $\mathbf{q}=(0.506, 0, -0.483)$ \cite{Forsyth1988,Ain1992}. A magnetically-induced ferroelectric polarization in this phase occurs in the $b$-direction \cite{Kimura2008}, which also exhibits ferroelectric hysteresis loops \cite{Kimura2008}. An electromagnon occurs in the multiferroic AF2 phase, which is active for oscillating electric fields parallel to the [101] direction and its strength is linked to the size of the static polarization \cite{Jones2014}. Between 229.3\,-\,230\,K an intermediate commensurate, collinear magnetic phase (AF3) forms \cite{Villarreal2012}, and above 230\,K is the paramagnetic phase (PM). The first-order nature of the AF1\,-\,AF2 phase transition has been observed by specific heat \cite{Junod1989,Loram1989,Gmelin1991} measurements.

The occurrence of the AF1\,-\,AF2 phase transition can be understood using the mechanism proposed by Yablonskii \cite{Yablonskii1990}, where a biquadratic exchange term stabilizes the low temperature AF1 state.  The free energy of a system is given by $F=E-TS$, where $T$ is temperature and $S$ is entropy. The energy of the interactions between the spin chains in the [10$\bar{1}$] direction is given by Yablonskii as \cite{Yablonskii1990}:

\begin{equation}
\begin{split}
E=\sum_{n}\big[J_{1}\mathbf{S}_{n}\cdot\mathbf{S}_{n+1}+J_{2}\mathbf{S}_{n}\cdot\mathbf{S}_{n+2} \\ +I(\mathbf{S}_{n}\cdot\mathbf{S}_{n+1})(\mathbf{S}_{n}\cdot\mathbf{S}_{n+2})\big],
\label{Yablonskii}
\end{split}
\end{equation}

\noindent
where $J_{1}$ is the nearest-neighbour ferromagnetic exchange interaction, $J_{2}$ is the next-nearest-neighbour antiferromagnetic exchange interaction, and $I$ is the biquadratic exchange interaction. Taking a mean-field approximation with the spatially averaged value of the spins as $S_{\mathrm{av}}$, the AF1 state will have a lower energy for $S_{\mathrm{av}}^{2}>J_{1}^{2}/8IJ_{2}$. As temperature increases, the value of $S_{\mathrm{av}}$ decreases, therefore at some point the AF2 state has a lower energy and the system undergoes a phase transition. The value of $S_{\mathrm{av}}$ is effectively controlled by disorder, which can be influenced in a number of ways, the simplest being a change in temperature changing the form of $F$ and driving the phase transition. Analogously a change in the entropy of the system will also change the form of $F$; evidence for this has been provided by optical-pump X-ray probe measurements on CuO \cite{Johnson2012}. With the system held just below the phase transition at 207\,K, electrons were excited above the charge-transfer bandgap by 800\,nm femtosecond pulses, introducing spin-disorder into the system and thus changing the entropy. A larger reduction in the intensity of the X-ray peak associated with the AF1 phase compared to the peak associated with the AF2 phase was observed, hinting at an ultrafast phase transition driven by the introduction of spin-disorder.

\section*{Quenched spin-disorder in CuO}

An alternative method of introducing spin-disorder into the system which may be more desirable for technological applications is by alloying with non-magnetic ions: theoretical investigations predict that the introduction of non-magnetic impurities may stabilize the multiferroic phase at higher temperatures than the pure case \cite{Giovannetti2011}, and that hydrostatic pressure can broaden the multiferroic phase above room temperature \cite{Rocquefelte2013}. Alloying of Cu$_{1-x}M_{x}$O with a non-magnetic ion $M$ has been demonstrated to broaden the multiferroic phase, for $M=$ Zn, Co by studying the static ferroelectric polarization along $b$ \cite{Hellsvik2014}, and for $M=$ Zn using the electromagnon response \cite{Jones2014a}. Spin-disorder introduced by the non-magnetic ions breaks up long-range correlations between spins and suppresses the N\'{e}el temperatures \cite{Hone1975}, reducing the AF1\,-\,AF2 transition to $\sim159$\,K and the AF2\,-paraelectric transition to $\sim190$\,K for $M=$ Zn and $x=0.05$ \cite{Jones2014a}. This equates to broadening the multiferroic phase from 17\,K to 30\,K. The electromagnon is preserved in the AF2 phase of Cu$_{0.95}$Zn$_{0.05}$O, with the same E\,$\parallel$\,[101] selection rule as in the pure case \cite{Jones2014,Jones2014a}.

The change in magnetic or lattice structure at a first-order phase transition is associated with many functional properties that emerge in the ordered phases \cite{Roy2013,Roy2014}. First-order phase transitions are characterized by a latent heat, due to a difference in entropy between the two phases requiring the system to absorb a fixed amount of energy in order for the transition to occur. This leads to phase coexistence, where some parts of the material have undergone the transition and others have not. In alloys, the dopant concentration can influence the nature of both the phase transition and the functional properties of interest \cite{Roy2013,Roy2014}. Imry and Wortis demonstrated that a first-order phase transition can be broadened by the introduction of quenched random disorder into the system \cite{Imry1979}. The correlation length of interactions can be reduced by local impurities, causing different parts of the material to undergo a first-order phase transition at different temperatures, appearing to round off the discontinuity in the order parameter or other related experimental observables.

Experimentally, a phase transition can be classified as first-order by the observation of a discontinuous change in the order parameter and a latent heat, phase coexistence, or hysteresis with respect to a control variable (such as temperature or external magnetic field). It is not necessary to observe hysteresis in the order parameter, as many experimental observables can be hysteretic over a first-order phase transition, exemplified by the observation of hysteresis in magneto-optical properties during the melting of a superconducting vortex-solid \cite{Soibel2000}. Broadening of first-order phase transitions by disorder may obscure discontinuities in experimental observables, making the latent heat of the transition difficult to determine \cite{Roy2014,Imry1979}. As such, hysteresis or phase coexistence are required to observe the first-order nature of the transition. 

Ideally, one should look for hysteresis in an experimental observable that is uniquely linked to the multiferroic state (e.g.\ the ferroelectric polarization), rather then observables that only change marginally across the transition (e.g.\ ultrasound velocity, magnetization). The pyroelectric current method is commonly used to study the polarization of magnetically-induced ferroelectrics: a large electric field is applied while in the multiferroic state, which is then removed and the pyroelectric current on cooling or heating through the ferroelectric/paraelectric phase transition is recorded and integrated to yield the polarization. This method often only measures the polarization as the temperature is swept out of the multiferroic state, and it is therefore challenging to study thermal hysteresis. 

As the electromagnon in Cu$_{1-x}$Zn$_{x}$O is experimentally observed via the dynamic excitation of the ferroelectric polarization, it cannot be observed without the AF2 phase being present within the material. The excitation energy and absorption strength of the electromagnon in CuO have previously been shown to closely track the size of the static polarization $P_{[010]}$ in the AF2 phase \cite{Jones2014}, which is in turn intimately linked to the magnetic order. Theoretical investigations of electromagnons in CuO indicate that the electromagnon energy is related to the size of the magnetoelectric coupling parameters in the spin Hamiltonian \cite{Cao2015}. Hence the properties of the electromagnon can provide a direct probe of the magnetic interactions giving rise to multiferroicity, and their evolution as the material undergoes a phase transition.

\section*{Experimental Methods}

\begin{figure*}[tb]
\includegraphics[width=1.0\textwidth]{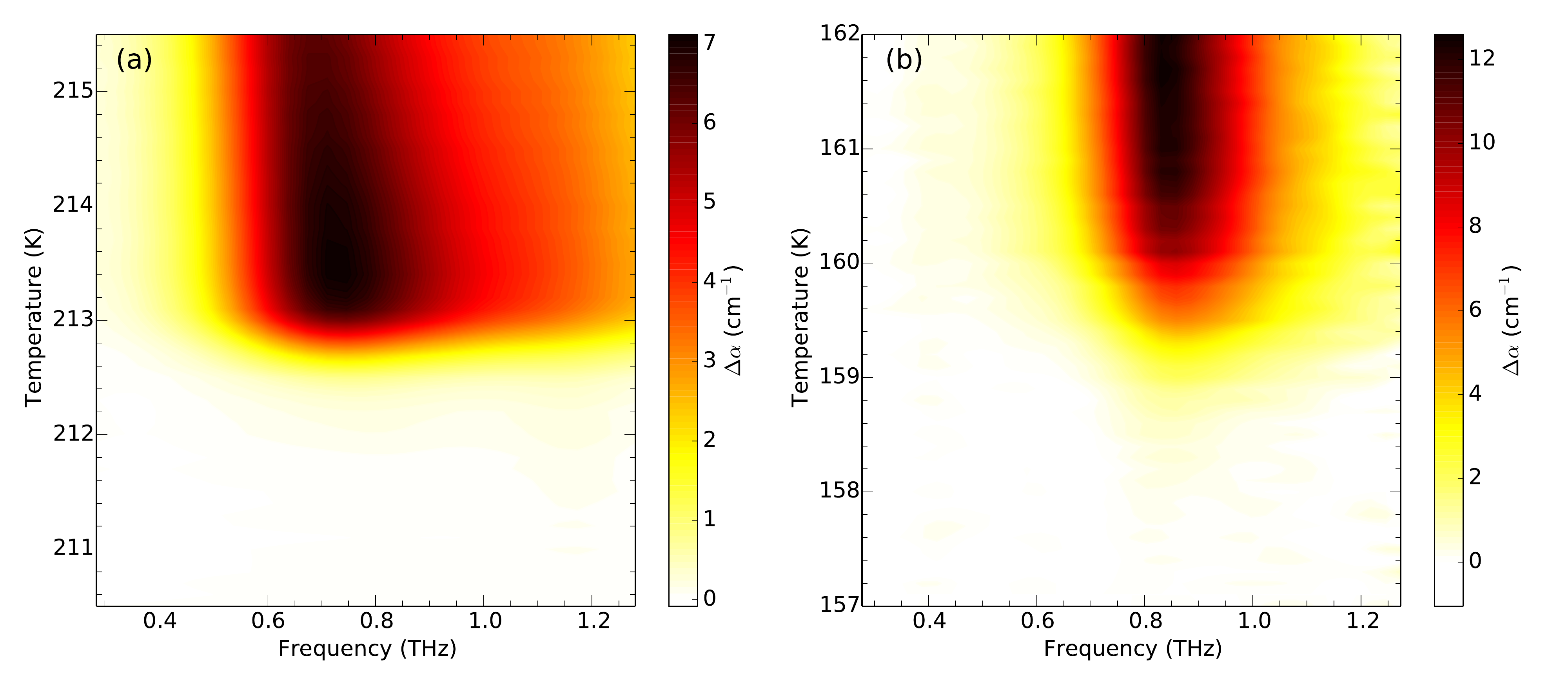}  
\caption{\small Change in terahertz absorption coefficient $\Delta\alpha$ with respect to the low temperature phase over the AF1\,-\,AF2 phase transition in Cu$_{1-x}$Zn$_{x}$O alloys, with \textbf{(a)} $x=0$ and \textbf{(b)} $x=0.05$.} 
\label{Figure1}
\end{figure*}

Single crystals of CuO and Cu$_{0.95}$Zn$_{0.05}$O were prepared by methods described elsewhere \cite{Prabhakaran2003,Jones2014a}. Samples of both materials were cut from the boule and aligned using Laue X-ray diffraction to have a ($10\bar{1}$) surface normal, giving the [101] and [010] crystallographic directions in the plane. Samples were double-side polished, resulting in a thickness of $1.28$\,mm for the CuO and $0.89$\,mm for the Cu$_{0.95}$Zn$_{0.05}$O.

The electromagnon response of the samples was measured by terahertz time-domain spectroscopy (THz-TDS) \cite{Jepsen2011,Lloyd-Hughes2012}. THz-TDS directly measures both the amplitude and the phase of the THz electric field after transmission through the sample, allowing the complex refractive index of the sample $\widetilde{n}=n+i\kappa$ to be determined. In order to avoid the influence of absorption from the broad $A^{3}_{u}$ phonon mode \cite{Jones2014a} or birefringent effects resulting from sample misalignment \cite{Jones2014,Mosley2017}, the relative change in terahertz absorption coefficient with respect to the low temperature AF1 phase $\Delta\alpha=\alpha(T)-\alpha(T_{\mathrm{AF1}})$, where $\alpha=2\omega\kappa/c$, was used instead of a vacuum reference.

Single-cycle, linearly-polarized pulses of THz radiation were generated using a wide-area interdigitated GaAs photoconductive emitter and detected via electro-optic sampling in ZnTe. The entire THz path inside the spectrometer was purged with dry N$_{2}$ gas in order to remove the influence of water vapour, which has strong absorption features at THz frequencies \cite{VanExter1989}. Samples were mounted in a cryostat such that the THz electric field was parallel to the [101] crystallographic direction, where the electromagnon absorption is strongest \cite{Jones2014,Mosley2017}, and positioned at the sample focus of the spectrometer. On heating and cooling the sample temperature was increased or decreased in discrete steps of 0.1\,K, with a wait time of 2 minutes before beginning data acquisition at each step, to ensure the sample was in thermal equilibrium with the cold finger of the cryostat. Data acquisition took 5 minutes at each step. The wait time and measurement time was the same for all scans on both samples.

\section*{Results and Discussion}

\subsection*{Electromagnon response over the AF1\,-\,AF2 phase transition}

To investigate the effects of zinc alloying on the AF1\,-\,AF2 phase transition in Cu$_{1-x}$Zn$_{x}$O, the change in terahertz absorption coefficient $\Delta\alpha$ over the AF1\,-\,AF2 phase transition is presented in Fig.\,\ref{Figure1}(a) for $x=0$ and Fig.\,\ref{Figure1}(b) for $x=0.05$. Samples were heated over a 5\,K range around the AF1\,-\,AF2 transition temperatures ($\sim213$\,K for $x=0$ and $\sim160$\,K for $x=0.05$), and $\Delta\alpha$ measured at each step. The electromagnon is observable in both samples as a peak in $\Delta\alpha$, at 0.72\,THz for $x=0$ and at 0.85\,THz for $x=0.05$, consistent with previous observations \cite{Jones2014,Jones2014a}. In the case of pure CuO, the electromagnon onset is rapid, with $\Delta\alpha$ rising from zero to a maximum in around 1.5\,K. Comparatively in the Zn-alloyed sample the onset is much slower, with the rise in absorption occurring in around 3.5\,K. At higher frequencies, a broad shoulder to the electromagnon around 1.2\,THz is also visible in the $x=0$ sample, and has a similar temperature-dependent onset as the main electromagnon. Intriguingly, this higher-frequency shoulder is not observed in the $x=0.05$ sample, suggesting that this mode may be either shifted in energy, suppressed or even disrupted entirely upon alloying with zinc.

\begin{figure}[t]
\includegraphics[width=0.5\textwidth]{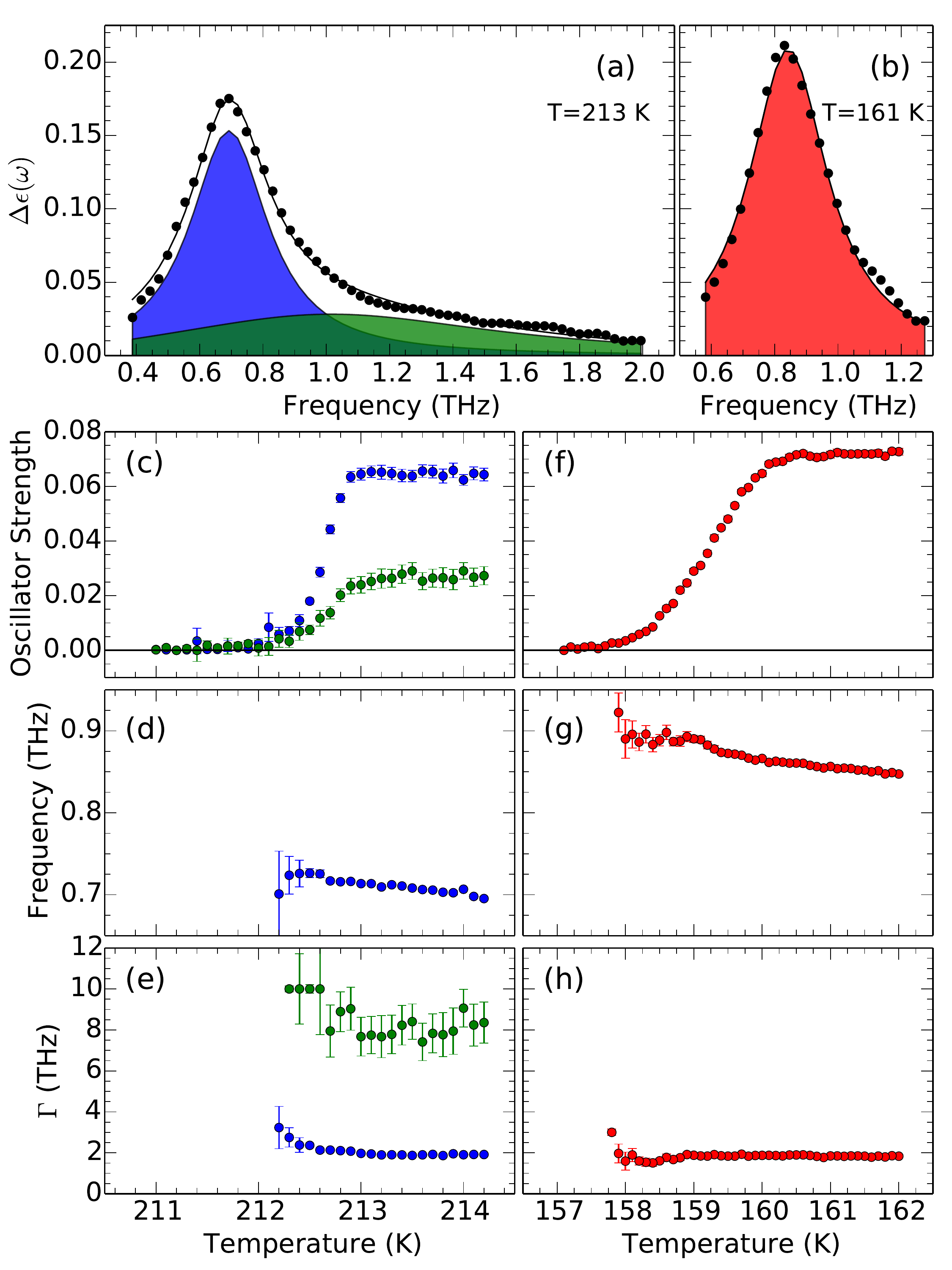}  
\caption{\small \textbf{(a)} and \textbf{(b)} are the temperature-induced change in dielectric function $\Delta\epsilon(\omega)$ (black dots) across the AF1\,-\,AF2 phase transition in Cu$_{1-x}$Zn$_{x}$O alloys with $x=0$ and $x=0.05$, respectively. Black lines are Drude-Lorentz fits comprised of a single oscillator (red shaded area) for $x=0.05$ and two oscillators (blue and green shaded areas) for $x=0$. \textbf{(c)}, \textbf{(d)} and \textbf{(e)} are the best fit parameters of the oscillator strength, frequency and linewidth $\Gamma$, respectively, over the phase transition for $x=0$. \textbf{(f)}, \textbf{(g)} and \textbf{(h)} are the corresponding best fit parameters over the phase transition for $x=0.05$. The color of the data points corresponds to the oscillator they refer to in \textbf{(a)} and \textbf{(b)}.} 
\label{Figure2}
\end{figure}

The electromagnon strength is intimately linked to the incommensurate magnetic order in the multiferroic phase, and therefore can give insight into the nature of the magnetic ordering. To evaluate the properties of the observed electromagnons in the Cu$_{1-x}$Zn$_{x}$O alloys, and to allow comparison with each other and others reported in the literature, a Drude-Lorentz oscillator model was used to fit the data. This gives the temperature-dependent change in dielectric function $\Delta\epsilon(\omega)=\epsilon(\omega,T_{2})-\epsilon(\omega,T_{1})$ as

\begin{equation}
\Delta\epsilon(\omega)=\frac{\Delta\epsilon\cdot\omega_{0}^{2}}{\omega^{2}_{0}-\omega^{2}-i\omega\Gamma},
\end{equation}

\noindent
where $\Delta\epsilon$ is the oscillator strength, $\omega_{0}$ is the oscillator frequency and $\Gamma$ is the linewidth. Fits to the experimentally measured change in dielectric function $\Delta\epsilon(\omega)$ of Cu$_{1-x}$Zn$_{x}$O with respect to the low-temperature phase are presented in Fig.\,\ref{Figure2}(a) for $x=0$ at 213\,K, and in Fig.\,\ref{Figure2}(b) for $x=0.05$ at 161\,K. A combination of two oscillators was required when fitting $\Delta\epsilon(\omega)$ for $x=0$, with the main electromagnon (blue) having a strength $\Delta\epsilon=0.065$, frequency $f=0.71$\,THz and linewidth $\Gamma=2.0$\,THz, while the broader shoulder mode (green) has $\Delta\epsilon=0.025$ and linewidth $\Gamma=8.0$\,THz. In order to correctly model the high-frequency features, the frequency of the shoulder mode was fixed at 1.2\,THz. Comparatively $\Delta\epsilon(\omega)$ for $x=0.05$ was well fit by a single oscillator (red), with strength $\Delta\epsilon=0.072$, frequency $f=0.84$\,THz and linewidth $\Gamma=2.0$\,THz. All values are consistent with those reported previously in the literature \cite{Jones2014,Jones2014a}. Experimental bandwidth was limited to 1.3\,THz in $x=0.05$ due to the softening and broadening of the phonon modes upon alloying with zinc \cite{Jones2014a}.

To quantify the evolution of the electromagnon across the phase transition in both samples, similar Drude-Lorentz fits to those in Fig.\,\ref{Figure2}(a) and \ref{Figure2}(b) were performed at each temperature step on cooling the samples over the AF1\,-\,AF2 phase transition. Fit parameters of oscillator strength, oscillator frequency and linewidth for $x=0$ are presented in Figs.\,\ref{Figure2}(c)\,-\,\ref{Figure2}(e). The strength of both oscillators begins increasing sharply at 212.2\,K, reaching a maximum in around 1.2\,K. The frequency of the main electromagnon mode continually red-shifts with increasing temperature over the phase transition, decreasing from 0.73\,THz at 212.2\,K to 0.69\,THz at 214.2\,K. The linewidth of both oscillators remains constant over the majority of the phase transition, decreasing slightly in width from when the excitation first begins to emerge. Fit parameters for $x=0.05$ are presented in Figs.\,\ref{Figure2}(f)\,-\,\ref{Figure2}(h). The strength of the oscillator shows a gentler, broader increase with temperature, starting to increase around 157.8\,K and reaching a maximum after around 3.5\,K. Both frequency and linewidth demonstrate the same trends seen for $x=0$, with the frequency red-shifting from 0.9\,THz at 167\,K to 0.85\,THz at 161\,K, and the linewidth remaining constant except for initially when the oscillator strength was very small. Values of frequency and linewidth are presented for temperatures at which the oscillator strength was large enough to provide a meaningful fit.

Theoretical investigations of electromagnons in CuO \cite{Cao2015} have modeled the magnetic properties using a spin Hamiltonian of the form

\begin{equation}
\begin{split}
\widehat{H}=\sum_{ij}J_{ij}\mathbf{S}_{i}\cdot\mathbf{S}_{j}+\mathbf{D}_{ij}\cdot(\mathbf{S}_{i}\times\mathbf{S}_{j}) \\ -\sum_{i}(\mathbf{K}\cdot\mathbf{S}_{i})^{2}+\widehat{H}_{\mathrm{me}},
\end{split}
\end{equation}

\noindent
where the first term accounts for superexchange interactions between adjacent spins $\mathbf{S}_{i}$, the second term decribes the DM interaction, the third term describes single-ion anisotropy, and the final term accounts for magnetoelectric coupling. Increases to the magnitude of $\mathbf{D}$ and $\mathbf{K}$, the DM and anisotropic coupling constants respectively, were found to enhance the electromagnon energy, directly linking the electromagnon energy to the strength of the interactions responsible for the AF2 phase. An electromagnon frequency of 0.73\,THz (energy $\sim3$\,meV) as observed at 212.3\,K in Fig.\,\ref{Figure2}(d) corresponds to $\left|\mathbf{D}\right|=0.4$\,meV and $\left|\mathbf{K}\right|=0.6$\,meV \cite{Cao2015}. The frequency of the electromagnon mode in Cu$_{1-x}$Zn$_{x}$O is observed to increase considerably upon alloying of $x=0.05$, suggesting that zinc alloying alters the interactions between spins. From reference \cite{Cao2015}, an increase in electromagnon frequency to 0.9\,THz (energy $\sim3.7$\,meV) as observed at 158.1\,K in Fig.\,\ref{Figure2}(g) corresponds to either an increase in $\left|\mathbf{D}\right|$ to 0.65\,meV, or an increase in $\left|\mathbf{K}\right|$ to 0.95\,meV, or a smaller simultaneous increase in both.

This alteration of magnetic interactions may be understood with reference to the ``order-by-disorder'' mechanism proposed by Henley \cite{Henley1989}, which describes the stabilization of a non-collinear spin state with respect to a collinear spin state, as quenched disorder favors spins in different sublattices to be oriented perpendicular to each other. Simulations have shown that the multiferroic phase in CuO can be stabilized by alloying with non-magnetic ions \cite{Hellsvik2014}. Density functional theory (DFT) calculations for pure CuO give a difference in energy between the AF2 and AF1 phase $\Delta{E}=E_{\mathrm{AF2}}-E_{\mathrm{AF1}}=2.15$\,meV per Cu, parameterised by $\Delta{E}=4IS^{4}$ where $S=1/2$ and $I$ is the biquadratic exchange coupling constant. Similar DFT calculations for alloyed CuO show a reduction in the energy difference between the AF1 and AF2 states, estimating a reduction to $\Delta{E}=1.94$\,meV per Cu for 5\% zinc alloying. Two contributions to this reduction in energy were determined, with similar sized contributions for Zn alloying: modification of the biquadratic exchange interactions between the non-magnetic impurity and spins on the other sublattice, and an alteration of the local Weiss field around the non-magnetic impurities which acts to orient spins on the same sublattice perpendicular to those on the other sublattice by the Henley mechanism. This change in the Weiss field around the non-magnetic ions will alter the value of the anisotropic coupling constant $\mathbf{K}$. A change in the DM coupling constant $\mathbf{D}$ can also occur as the DM interaction is sensitive to the Cu-O-Cu bond angle \cite{Katsura2005}, which changes on alloying with zinc \cite{Jones2014}.

\subsection*{Disorder-broadening of the AF1\,-\,AF2 phase transition}

\begin{figure}[t]
\includegraphics[width=0.5\textwidth]{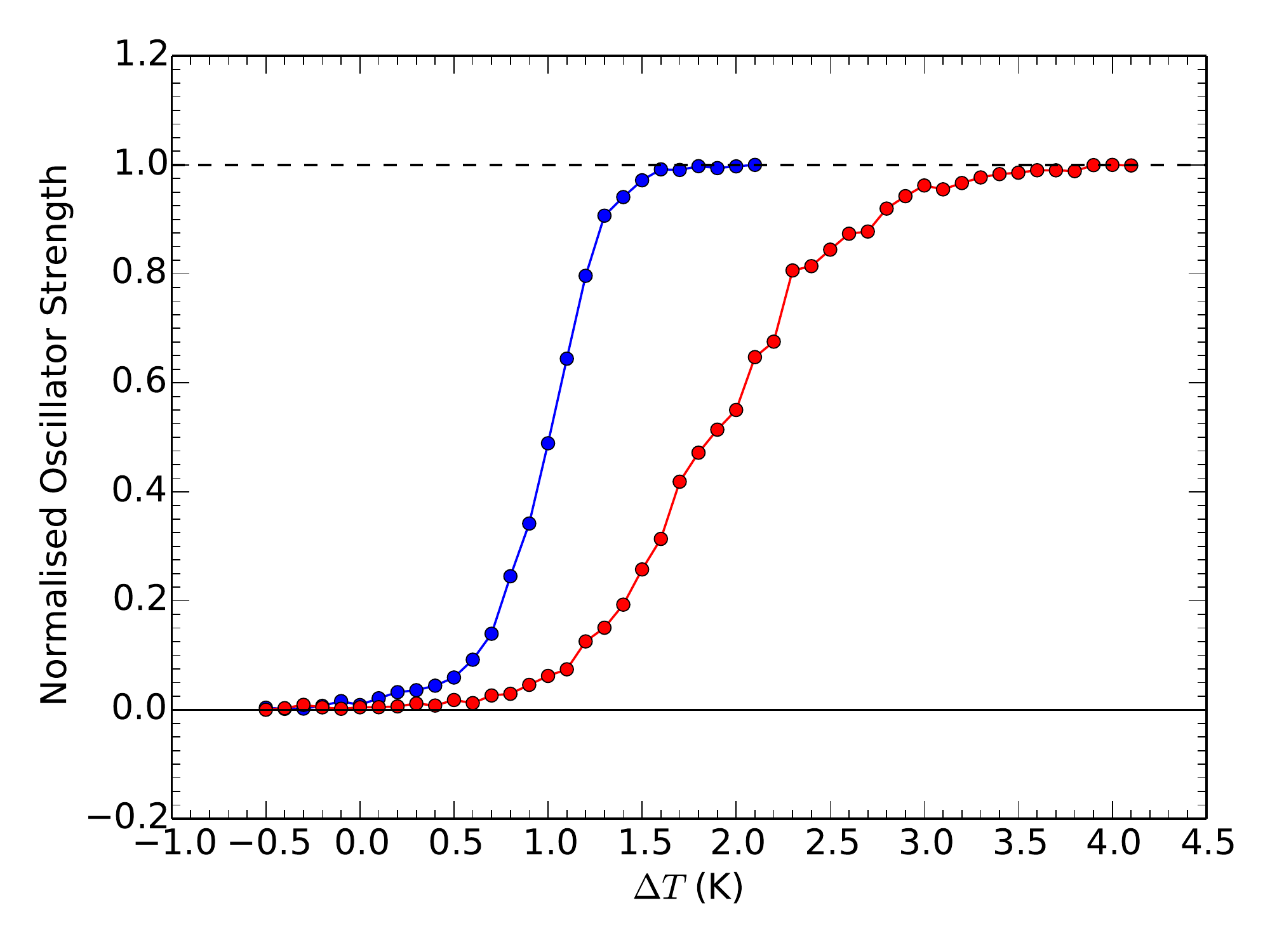}  
\caption{\small Temperature dependent change in the oscillator strength of the electromagnon, over the AF1\,-\,AF2 phase transition, in Cu$_{1-x}$Zn$_{x}$O alloys with $x=0$ (blue points) and $x=0.05$ (red points).} 
\label{Figure3}
\end{figure}

For a direct comparison of the width of the AF1\,-\,AF2 phase transition for $x=0$ and $x=0.05$, Fig.\,\ref{Figure3} presents the oscillator strength of the two main modes from Figs.\,\ref{Figure1}(a) and \ref{Figure1}(b), normalized by their maximum value. The change in temperature $\Delta{T}$ is defined relative to the temperature at which the derivative of the oscillator strength with respect to temperature, $d(\Delta\epsilon)/dT$, increases from zero. The oscillator strength rises from zero to maximum in around $\Delta{T}_{1}=1.6$\,K for $x=0$ and $\Delta{T}_{2}=3.9$\,K for $x=0.05$. Thus a broadening of 2.3\,K occurs for the AF1\,-\,AF2 phase transition upon alloying with 5\% zinc, a ratio of $\Delta{T}_{2}/\Delta{T}_{1}=2.6$.

This broadening of the phase transition occurs due to the quenched random spin-disorder introduced upon alloying with non-magnetic zinc ions. As mentioned previously, CuO can be regarded as a quasi-1D colinear Heisenberg antiferromagnet in the commensurate AF1 phase, consisting of spin chains along the [10$\bar{1}$] direction. Alloying of $x=0.05$ has been shown to suppress the N\'{e}el temperatures in Cu$_{1-x}$Zn$_{x}$O \cite{Jones2014a} due to spin disorder breaking communication along the spin chains, which reduces the correlation length to the average impurity separation. During formation of the alloy, the local impurity density in the melt will vary slightly around an average value, and this variation becomes frozen-in as the alloy crystallizes. The correlation length of interactions therefore varies depending on the local impurity density, and hence so does the local N\'{e}el temperature. In local regions where the impurity density is higher than average the N\'{e}el temperature will be lower than average, and vice versa. This causes the AF1\,-\,AF2 phase transition to occur over a range of temperatures depending on the local impurity density \cite{Imry1979}.

The electromagnon excitation cannot be observed without the presence of the AF2 phase, as its strength is strongly linked to the size of the magnetically-induced static polarization in the AF2 phase \cite{Jones2014}; as such the normalised oscillator strength in Fig.\,\ref{Figure3} can be interpreted as the ratio of the amount of AF2 to AF1 phase present in the sample. The emergence of the electromagnon upon heating the sample through the phase transition therefore corresponds to initial nucleation of the AF2 phase. The local variation in N\'{e}el temperature resulting from quenched spin-disorder in the $x=0.05$ sample causes nucleation to occur at different temperatures in different regions of the sample, evidenced by the broader increase in electromagnon strength with temperature.

\subsection*{Hysteresis in the electromagnon response}

\begin{figure}[t]
\includegraphics[width=0.5\textwidth]{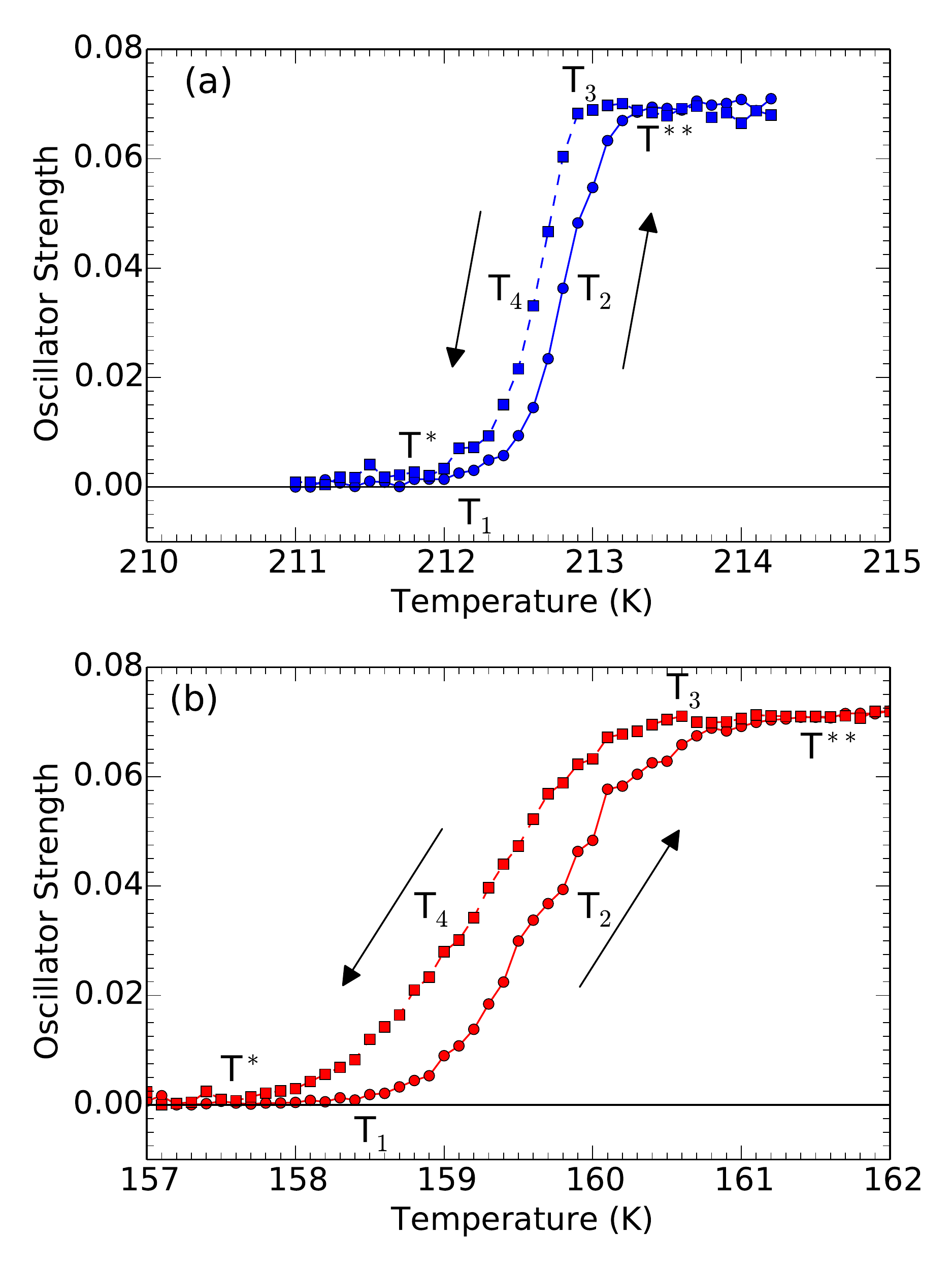}  
\caption{\small Temperature-dependent hysteresis observed in the electromagnon response across the AF1\,-\,AF2 phase transition in Cu$_{1-x}$Zn$_{x}$O alloys with \textbf{(a)} $x=0$ and \textbf{(b)} $x=0.05$. Upwards facing arrows denote increasing temperature (also denoted by solid circles) and downwards facing arrows denote decreasing temperature (solid squares). $T_{1}$ and $T_{3}$ are the temperatures at which the oscillator strength begins to deviate significantly from zero or the maximum, respectively. $T_{2}$ and $T_{4}$ are the mid-points of the phase transition during heating and cooling, respectively. $T^{**}$ and $T^{*}$ are the limits of metastability for superheating and supercooling, respectively.} 
\label{Figure4}
\end{figure}

Temperature-dependent hysteresis is one of the experimental signatures of a first-order phase transition, and should be present in many experimental observables related to an order parameter involved in the phase transition. As a natural extention to the analysis of Fig.\,\ref{Figure3}, we show here that the strength of the electromagnon can be used to observe hysteretic behaviour at the AF1\,-\,AF2 phase transition on heating and cooling of the sample.

The electromagnon response in Cu$_{1-x}$Zn$_{x}$O alloys as the temperature is first increased and then decreased across the AF1\,-\,AF2 phase transition is presented in Fig.\,\ref{Figure4}(a) for $x=0$ and in Fig.\,\ref{Figure4}(b) for $x=0.05$. Hysteresis is clearly visible in both pure and zinc alloyed samples, with the centre of the phase transition occurring at higher temperatures when temperature was increased over the phase transition compared to when it was decreased. The centre points of the increasing and decreasing temperature measurements, $T_{2}$ and $T_{4}$ respectively, are separated by 0.2\,K in the $x=0$ sample and 0.5\,K in the $x=0.05$ sample. The ratio between these values of 2.5 is similar to the degree of disorder-induced broadening of the phase transition observed in Fig.\,\ref{Figure3}. Subsequent heating and cooling measurements reproduced the same hysteretic behaviour.

\begin{figure}[t]
\includegraphics[width=0.5\textwidth]{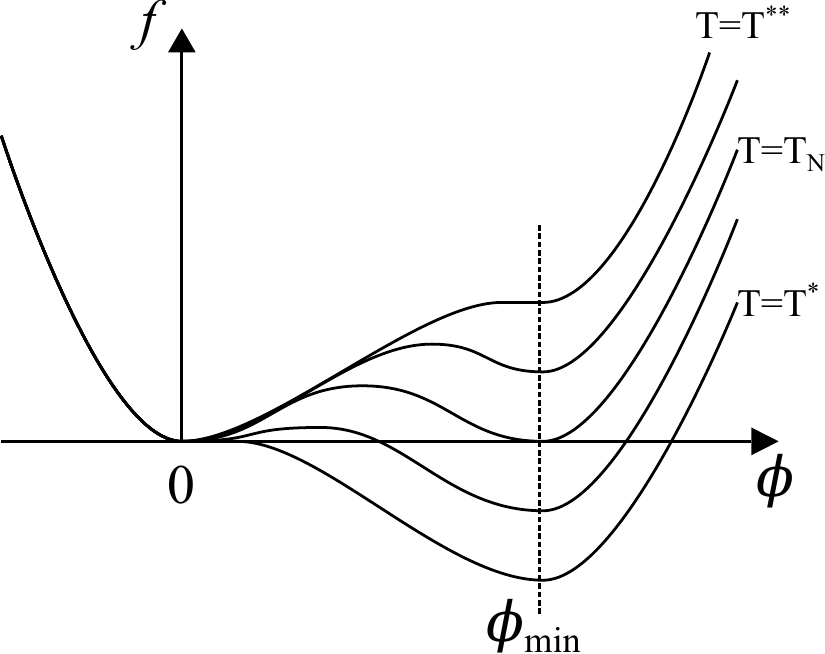}  
\caption{\small Schematic diagram of the free energy $f$ as temperature is increased and decreased across a first-order phase transition, with respect to the order parameter $\phi$. $T_{N}$ is the thermodynamic transition temperature, $T^{*}$ is the metastable limit of supercooling, and $T^{**}$ is the metastable limit of superheating.} 
\label{Figure5}
\end{figure}

Here we consider a simple phenomenological description of the free energy with reference to Landau theory for a first-order phase transition, in order to describe the observed hysteretic behaviour. Advanced treatments more suitable for multiferroics have been reported, for instance including the magnetoelectric coupling between a uniform polarization and incommensurate magnetic order \cite{Harris2007}, or non-local Landau theories that describe spin-lattice coupling \cite{Plumer2008,Villarreal2012}. The free energy density of the system with respect to a control variable, here taken to be the temperature $T$, can be expressed in terms of the order parameter $\phi$. One possibility by which a first-order phase transition can occur is if the free energy density has a third-order term in $\phi$, such as

\begin{equation}
f(T,\phi)=\frac{r(T)}{2}\phi^{2}-w\phi^{3}+u\phi^{4},
\end{equation}

\noindent
where $w$ and $u$ are positive and temperature-independent. In many systems third-order terms in $\phi$ are forbidden by symmetry. In such a case, a first-order phase transition can still occur if the free energy has a negative quartic term with a positive sixth-order term for stability, such as

\begin{equation}
f(T,\phi)=\frac{r(T)}{2}\phi^{2}-w\phi^{4}+u\phi^{6}.
\label{FOLandau}
\end{equation}

\noindent Fourth- and sixth-order terms can arise in magnetoelectrics, for instance as a result of strong spin-lattice coupling \cite{Plumer2008}, and such Landau models have been developed for CuO \cite{Villarreal2012,Quirion2013}. Substantial spin-lattice coupling has been reported in Cu$_{1-x}$Zn$_x$O as the phonon modes alter across the AF1-AF2 transition \cite{Jones2014a}. Note that the order parameter can be complex, to allow for non-collinear magnetic states \cite{Quirion2013}.

A schematic diagram of the form of the free energy density around a first-order phase transition with respect to $\phi$ is presented in Fig.\,\ref{Figure5}. Here we assume the order parameter for CuO to be represented by the magnetisation along $b$ for one magnetic sublattice, which is zero in the spin-cycloid phase (AF2) and a finite value $\phi_{\mathrm{min}}$ in the antiferromagnetic AF1 phase. When the temperature is higher than the thermodynamic transition temperature $T_{N}$ the free energy has one global minimum, at $\phi=0$, corresponding to the AF2 phase. At $T=T_{N}$ there are two minima to the free energy and the AF1 and AF2 phases can coexist. These minima are located at $\phi=0$ and $\phi=\phi_{\mathrm{min}}$ and separated by an energy barrier. When the temperature drops below $T_{N}$ the global minimum of the system shifts to $\phi_{\mathrm{min}}$, corresponding to the AF1 phase; however $\phi=0$ remains a local minimum of the system, with the two phases still separated by an energy barrier, which will reduce in height with decreasing temperature. Eventually the height of the energy barrier will reduce to zero at the limit of metastability for supercooling $T=T^{*}$, the temperature at which $d^{2}f/d\phi^{2}=0$ at $\phi=0$ and none of the AF2 phase can exist. An analogous process occurs for increasing temperature, where $\phi_{\mathrm{min}}$ continues to be a local minimum of the system until the limit of metastability for superheating $T=T^{**}$, the temperature at which $d^{2}f/d\phi^{2}=0$ at $\phi_{\mathrm{min}}=0$ and none of the AF1 phase can exist. It is these superheating and supercooling effects that give rise to the hysteresis in experimental observables with respect to the control parameters.

The hysteretic evolution of the oscillator strength with temperature for both pure and zinc alloyed samples can be explained with reference to the free energy diagram in Fig.\,\ref{Figure5}, and the nomenclature used to describe hysteresis of disorder-broadened first-order phase transitions outlined in reference \cite{Roy2014}. In the case of an ideally pure compound and in the absence of temperature fluctuations, the high or low temperature states can be taken to the limit of metastability at $T^{*}$ and $T^{**}$ respectively, where the phase transition occurs as a sharp discontinuity in the hysteretic parameter being observed.

In any real system, temperature fluctuations will perturb the system and destroy the metastable state at some intermediate temperature between $T_{N}$ and $T^{*}$ or $T^{**}$. The small amount of disorder present in any real crystal from impurities or defects will also cause a slight rounding of the transition. Both of these effects are observable in the hysteresis of pure CuO in Fig.\,\ref{Figure4}(a). On heating, the electromagnon excitation emerges at temperature $T_{1}$, signifying the onset of nucleation of the AF2 phase, and grows in strength until reaching a maximum at 213.4\,K. On cooling, the electromagnon strength begins to differ from that observed during heating around 213.4\,K, signifying this temperature as $T^{**}$, and begins to decrease in strength at temperature $T_{3}$ signifying the onset of nucleation of the AF1 phase. The temperature at which the electromagnon strength reduces to zero on cooling can therefore be identified as $T^{*}$. The system is observed to be in the metastable state for superheating and supercooling over a temperature range of 0.4\,K in both cases. Similar trends can be seen for the zinc alloyed sample in Fig.\,\ref{Figure4}(b), however in this case the phase transition is broadened by spin-disorder. The metastable phases for superheating and supercooling are also broadened upon zinc alloying, with both increasing to a width of 1\,K.

As discussed previously, simulations of the magnetic interactions in alloyed CuO showed a decrease in the energy difference between the AF1 and AF2 states relative to pure CuO \cite{Hellsvik2014}. However, the metastable phase observed in the zinc alloyed sample is wider than the pure sample. This may be partially due to smaller thermal fluctuations at the lower N\'{e}el temperatures of the alloyed sample, but could also be due to modifications to the exchange coupling parameters on alloying altering the form of the free energy barrier between the two phases. Modification of the biquadratic exchange and higher order coupling terms by the non-magnetic ions, as discussed in terms of the ``order-by-disorder" mechanism above, could therefore possibly alter the free energy barrier height and increase the stability of the metastable phases in the zinc alloyed sample.

\section*{Conclusion}

Here we reported for the first time that the dynamic magnetoelectric response at terahertz frequencies can be used to probe hysteresis and spin-disorder broadening at magnetic phase transitions. The oscillator strength $\Delta\epsilon$ of the electromagnon in Cu$_{1-x}$Zn$_x$O was taken to represent the fraction of the magnetically-induced ferroelectric phase, AF2. By precisely tracking $\Delta\epsilon$ on cooling or heating across the phase transition between the commensurate AF1 phase and the ferroelectric AF2 phase we observed thermal hysteresis, indicating the first-order nature of this transition. The limits of metastability for superheating and supercooling were identified, and the metastable region was found to broaden for the alloy with greater spin-disorder. Alloying also enhanced the electromagnon energy from 3.0\,meV to 3.7\,meV, potentially as a result of enhanced DM interaction or greater single-ion anisotropy. The results were discussed within the context of a simple Landau theory of a first-order phase transition, with reference to recent advances in phenomenological and first-principles theoretical models of CuO.

This work may have immediate impact by providing a new way to study the nature of magnetic phase transitions in multiferroics. In the longer term the increased understanding of multiferroics yielded by ultrafast spectroscopic methods, including terahertz time-domain spectroscopy, may help develop new magnetoelectric and multiferroic materials for spintronic applications.

\section*{References}


\section*{Acknowledgements}
The authors acknowledge funding from the EPSRC (UK). Data related to this publication is available from the University of Warwick data archive at http://wrap.warwick.ac.uk/89XXX.

\end{document}